\documentclass[aps,reprint,showpacs,amsmath,amssymb,prc,floatfix,nofootinbib]{revtex4-1}
\usepackage{color}
\usepackage{graphicx}
\usepackage{dcolumn}
\usepackage{bm}
\usepackage{rotating}
\usepackage{float}
\usepackage{xcolor}
\usepackage{natbib}
\usepackage{color}
\usepackage{times}
\usepackage[utf8x]{inputenc}
\usepackage[colorlinks,citecolor=blue,linkcolor=red,anchorcolor=blue,filecolor=blue,urlcolor=blue]{hyperref}
\usepackage{comment}
\begin{document}
\title{Isotopic shift and search of magic number in the superheavy region}

\author{Jeet Amrit Pattnaik$^{1}$}
\email{jeetamritboudh@gmail.com}
\author{R. N. Panda$^{1}$}
\email{rabinarayanpanda@soa.ac.in}
\author{M. Bhuyan$^{2,3}$}
\email{bunuphy@um.edu.my}
\author{S. K. Patra$^{4,5}$}
\email{patra@iopb.res.in}

\affiliation{$^1$Department of Physics, Siksha $'O'$ Anusandhan, Deemed to be University, Bhubaneswar-751030, India}
\affiliation{$^2$Center of theoretical and Computational Physics, Department of Physics, University of Malaya, Kuala Lumpur, 50603, Malaysia}
\affiliation{$^3$Institute of Research and Development, Duy Tan University, Da Nang 550000, Vietnam}
\affiliation{$^4$Institute of Physics, Sachivalya Marg, Bhubaneswar-751005, India}
\affiliation{$^5$Homi Bhabha National Institute, Training School Complex, Anushakti Nagar, Mumbai 400094, India}
\date{\today}
\begin{abstract}
The ground state bulk properties such as binding energy, root-mean-square radius, pairing energy, nuclear density distributions, and single-particle energies are calculated for the isotopic chain of Ca, Sn, Pb, and Z = 120 nuclei. The relativistic mean-field with recently developed G3, IOPB-1, and Relativistic-Hartree-Bogoliubov with density-dependent DD-ME1 and DD-ME2 parameter sets are used in the present analysis.  The respective shifts over the isotopic chain for the structural observables and surface property like symmetry energy are also estimated using a three-point method, which is crucial for the systematic analysis of the shell/sub-shell closure. The calculated results are compared with the available experimental data for various bulk properties, wherever available.  A multiple isotopic shift leads to the shell/sub-shell closure at N = (20 \& 28), (50 \& 82), and 126 for Ca, Sn, and Pb isotopes, respectively, are observed.  The analysis also supports the neutron magic at N = 40 and 184 for highly neutron-rich $^{60}$Ca, and $^{304}$120, predicted to be the next double magic beyond $^{208}$Pb, respectively. Observing the occupancy number, we notice the higher neutron orbitals are mostly occupied before the lower one, which causes the kinks at neutron magic with an amalgam in the isotopic chain trend above nuclei. We also notice the correlation between the occupation probabilities and the magicity of a nucleus and vice-versa.
\end{abstract}
\maketitle
\section{Introduction}
\label{sec1}
The nuclear isotopic shift is one of the benchmarks for studying the nuclear interaction \cite{mack56,neumann76}. Although the older Skyrme interactions reproduce the nuclear matter and finite nuclear properties pretty well, these forces fail to produce the isotopic shift of Pb nuclei contrary to the experimental observations \cite{chabanat97,stone07}. After the success of the relativistic mean-field (RMF) theory of reproducing the bulk properties of finite nuclei, Sharma {\it et al.} \cite{sharma} tested the RMF results as input to analyze the isotopic shift of charge radius for Pb nuclei. The calculated results match well with the experimental trend. However, these analyses were confined to N=126 only for a long time. Recently, Goddard {\it et al.} \cite{riosprl} revisited the isotopic shift at N = 126 for Pb-isotopes using advanced Skyrme Energy Density Functional, and Gorges {\it et al.} \cite{georges2019}, reported that the neutron number N = 82 also shows the isotopic shift for Sn nuclei. In these works, the charge radius, pairing, surface properties, single-particle energies, occupation, and spin-orbit couplings are used for the isotopic shift indicators. The results are interesting for the present scenario of nuclear structure studies to fill the valence orbitals. Further, the charge distribution is dominated by quantum shell effects for a shell/sub-shell closure nuclei/isotope, and that causes an irregularity in the systematic trend, so-called the isotopic shift. In other words, the nuclei having properties large shell gap in single-particle energy levels, doubly magic, and with zero pairing energy always show the kinks and peaks in their anomalous trend \cite{riosepj,georges2019,bhujpg2021}. Isotopic shifts are the gateway to achieve the shell closure because they appear only when a large shell gap is maintained amongst the levels. The added nucleons attend higher orbital, resulting in an energy difference \cite{webb}. Thus isotopic shift contributes a trivial aspect in finding the magicity/shell closure in the nuclear chart and vice-versa.

The isospin dependence of neutron-skin thickness connected to the symmetry energy and its slope parameter of infinite nuclear matter has now been focused as one of the primary research subjects in nuclear physics. New generation RI beam facility, namely, RIBF at RIKEN in Japan \cite{moto10}, FAIR at GSI in Germany \cite{aug07}, GANIL in France \cite{gale10}, and FRIB in the USA \cite{thoe10} have been and/or will be operated with preliminary results to explain the exotic nature of drip-line nuclei. Here, we accorded brief information for the Ca, Sn, and Pb region of the nuclear chart. From neutron number N=54 to N=64, these regions are constantly being exceptional in the nuclear landscape for Sn nuclei \cite{nayak99}. Earlier in 2004, Satapathy {\it et al.} \cite{patrajpg} explained the above region precisely in their work using the infinite nuclear matter model. One more specialty of this region is their deformed ground state, and higher shape degrees of freedom \cite{oulne19, patrajpg}. They are always showing very uncommon pairing, chemical potential energy, deformation parameter, etc., and also very stable enough, just like magic numbers. So there is a possibility of getting kinks or shifts in this region. Satapathy {\it et al.} \cite{patrajpg} have also been predicted Z = 62 is considered to be a deformed magic number. Most of the theoretical approaches will fail to explain the nuclear matter characteristics of the Sn region \cite{subratsoftness} and reference therein. So this is the beauty of the Sn isotopic chain. Like the Sn nucleus, Ca also has a neutron number which is behaving like magic numbers. At neutron number N = 40, it is noticed to be a sub-shell closure for the Ca nucleus. While we are talking about the Pb region, one should know that it is a spotless region. There is no discrepancy happening in that region. This is all about the known nuclei regions, but what about the unknown magic shell nuclei Z=120 isotopic chain. We are very much enthusiastic to know the characteristics of the superheavy region.

The motivation of the present manuscript is to see the isotopic shifts for the magic nuclei Ca, Sn, Pb, and Z = 120 (predicted)). Here, binding energy (BE), charge radius ($R_{ch}$), two neutron separation energy ($S_{2n}$), neutron skin thickness $(\triangle{R_{np}})$, neutron chemical energy ($\lambda_n$) and symmetric energy ($S^A$) are used as assets to observe the shifts. Studying the characteristics of famous Z=120 nuclei has always been an interesting matter to discuss. Many people have reported about the shell closure at Z = 120 \cite{sharmalala,bender,khoo,prassa}, where they have also mentioned about the corresponding neutron numbers N=172 and 184 \cite{guptapatra,webb,tsil,tani20}. However, we are curious to see how these neutron numbers will react to the isotopic shifts in the superheavy region. Most of the researchers have made a clear vision about the combination, i.e., Z=120 with N=184, the next magic in the superheavy valley by considering the factors like pairing energy, chemical potential energy, deformation parameter, single-particle spectra, and decay energies \cite{agbemava,shi,bhu12,subratijmpe}. We have obtained the shifts for the charge radius and other bulk properties, surface property like symmetry energy in this work. Further, it is interesting to examine the isotopic shift for Z =120 at the predicted neutron magic number N = 184. The relativistic mean-field (RMF) with G3 \cite{kumar17,kumar18}, IOPB-I \cite{kumar18}, density dependant DD-ME1 \cite{pring2002}, and DD-ME2 \cite{pring2005} are used in the present analysis. It is worth mentioning that these parameters successfully reproduce the experimental quantities for almost all known nuclei \cite{kumar17,kumar18, pring2002, pring2005}.

The paper is organized as follows: the relativistic mean-field model and the fitting procedure of energy density functional ${\cal{E}(\rho)}$ to an analytical expression in coordinate space with the help of the Br\"uckner prescription \cite{bruk68,bruk69} are discussed in Sec. \ref{theory}. The coherent density fluctuation model (CDFM) is also briefly outlined in this Section. The calculated results are discussed in Sec. \ref{results}. A summary and concluding remarks are presented in Sec. \ref{summary}. \\

\section{Relativistic mean field (RMF) Model}
\label{theory}
The standard nonlinear RMF Lagrangian density is built up by the interactions of isoscalar-scalar $\sigma$, isoscalar-vector $\omega$, and isovector-vector $\rho$ mesons with nucleons. In this extended version, the Lagrangian includes the self and cross-coupling of the $\sigma$, $\omega$, and $\rho$ mesons, entitled the effective field theory motivated relativistic mean-field formalism \cite{frun96,frun97,kumar17,kumar18}. All the possible meson-nucleon and their self-interactions and interactions of $\delta$ mesons are considered in these models. The cross-coupling of $\rho$ mesons with $\sigma$ and $\omega$ mesons allows for varying neutron skin thickness in heavy mass nucleus like $^{208}$Pb, and also the self couplings of $\sigma$ mesons reduces the nuclear matter incompressibility \cite{tuhin,kumar17}. For a nucleon-meson interacting system, the energy density functional in terms of Lagrangian can be given as \cite{kumar18}:
\begin{widetext}
\begin{eqnarray}
{\cal E}({r})&=&\sum_{\alpha=p,n} \varphi_\alpha^\dagger({r})\Bigg\{-i \mbox{\boldmath$\alpha$} \!\cdot\!\mbox{\boldmath$\nabla$}+\beta \bigg[M-\Phi (r)-\tau_3 D(r)\bigg]+ W({r})+\frac{1}{2}\tau_3 R({r})+\frac{1+\tau_3}{2} A({r})-\frac{i\beta\mbox{\boldmath$\alpha$}}{2M}\!\cdot\!\bigg(f_\omega\mbox{\boldmath$\nabla$}W({r})
\nonumber\\
&&
+\frac{1}{2}f_\rho\tau_3 \mbox{\boldmath$\nabla$}R({r})\bigg)\Bigg\} \varphi_\alpha(r)+\left(\frac{1}{2}
+\frac{\kappa_3}{3!}\frac{\Phi({r})}{M}+\frac{\kappa_4}{4!}\frac{\Phi^2({r})}{M^2}\right)
\frac{m_s^2}{g_s^2}\Phi^2({r})-\frac{\zeta_0}{4!}\frac{1}{g_\omega^2 }W^4({r})+\frac{1}{2g_s^2}\left(1+\alpha_1\frac{\Phi({r})}{M}\right) \bigg(
\mbox{\boldmath $\nabla$}\Phi({r})\bigg)^2
\nonumber\\
&&
-\frac{1}{2g_\omega^2}\left( 1 +\alpha_2\frac{\Phi({r})}{M}\right)\bigg(\mbox{\boldmath$\nabla$} W({r})\bigg)^2-\frac{1}{2}\left(1+\eta_1\frac{\Phi({r})}{M}+\frac{\eta_2}{2} \frac{\Phi^2({r})}{M^2}\right)\frac{m_\omega^2}{g_\omega^2} W^2({r})-\frac{1}{2e^2} \bigg( \mbox{\boldmath $\nabla$} A({r})\bigg)^2-\frac{1}{2g_\rho^2} \bigg( \mbox{\boldmath $\nabla$} R({r})\bigg)^2
\nonumber\\
&& 
-\frac{1}{2} \left( 1 + \eta_\rho \frac{\Phi({r})}{M}\right)\frac{m_\rho^2}{g_\rho^2} R^2({r}) -\Lambda_{\omega}\bigg(R^{2}(r)\times W^{2}(r)\bigg)+\frac{1}{2 g_{\delta}^{2}}\left( \mbox{\boldmath $\nabla$} D({r})\right)^2+\frac{1}{2}\frac{{m_{\delta}}^2}{g_{\delta}^{2}}D^{2}(r)\;.
\label{eds}
\end{eqnarray}
\end{widetext}
Here $\Phi$, $W$, $R$ and $D$ are the redefined fields $\Phi = g_s\sigma$, $W = g_\omega \omega$, $R$ = g$_\rho\vec{\rho}$ and $D=g_\delta\delta$. The coupling constants $g_\sigma$, $g_\omega$, $g_\rho$, $g_\delta$, and the masses $m_\sigma$, $m_\omega$, $m_\rho$ and $m_\delta$ respectively given for $\sigma$, $\omega$, $\rho$, and  $\delta$ mesons. $\frac{e^2}{4\pi}$ is the photon coupling constant.From Eq. (\ref{eds}), we obtain our energy density ${\cal{E}}_{nucl.}$  G3 \cite{kumar17}, IOPB-I \cite{kumar18}, DD-ME1 \cite{pring2002}, DD-ME2 \cite{pring2005} by taking that the exchange of mesons establish a uniform field, where the oscillations done by nucleons in a periodic motion said to be simple harmonic. From the effective-RMF energy density, the equation of motions (EoS) for the mesons and the nucleons is procured using the Euler-Lagrange equation. A bunch of coupled differential equations is retained and settled accordingly \cite{kumar18}. The scalar and vector densities,
\begin{eqnarray}
\rho_s(r)&=&\sum_\alpha \varphi_\alpha^\dagger({r})\beta\varphi_\alpha, \label{scaden}\\
\end{eqnarray}
and
\begin{eqnarray}
\rho_v(r)&=&\sum_\alpha \varphi_\alpha^\dagger({r})\tau_{3}\varphi_\alpha\label{vecden},
\end{eqnarray}
respectively, are figured out from the converged elucidations within the spherical harmonics.  In the account of a detailed study of the bulk properties of close-shell nuclei in the super-heavy region, we will improvise our energy density. With the help of that, we will produce our required quantities binding energy (B.E), charge radii, and root-mean-square radii, etc. The spherical densities are used within the CDFM to find the weight function $|F(x)|^2$, which is a crucial quantity to obtain the symmetry energy ($S^{A}$). \\

\subsection{Coherent Density Fluctuations Model Formalism} 
\label{CDFM}
The Coherent Density Fluctuations Model (CDFM) is firstly prescribed by Antonov {\it et al.} \cite{antozphys1980}, which plays a vital role in taking care of the fluctuation of momentum and coordinate. This method can be easily used to interpret the surface property like symmetry energy of finite nuclei. In this CDFM formalism, we can use NM tool $S^{NM}$ from Eqs. (\ref{s0}) to obtain their values for a finite nucleus \cite{anto4,anto2,anto3,antozphys1980}. Within this model, the density  $\rho$ ({\bf  r, r$'$}) of a finite nucleus can be rewritten as the coherent superposition of infinite number of one-body density matrix (OBDM)  $\rho_x$ ({\bf  r}, {\bf  r$'$}) for spherical parts of NM specified as {\it  Fluctons} \cite{bhu18,gad11},
\begin{equation}
\rho_x ({\bf  r}) = \rho_0 (x)\, \Theta (x - \vert {\bf  r} \vert),
\label{denx}
\end{equation}
with $\rho_o (x) = \frac{3A}{4 \pi x^3}$. The generator coordinate x is the radius of a sphere consisting of Fermi gas having all the  A nucleons distributed uniformly over it. It is appropriate to demonstrate for such a system the OBDM disclosed as below \cite{bhu18,anto2,gad11,gad12},
\begin{equation}
\rho ({\bf  r}, {\bf  r'}) = \int_0^{\infty} dx \vert F(x) \vert^2 \rho_x ({\bf  r}, {\bf  r'}).
\label{denr}
\end{equation}
Here $\vert F(x) \vert^2 $ is called as weight function (WF). The coherent superposition of OBDM $\rho_x ({\bf  r}, {\bf  r'})$ is given below as:
\begin{eqnarray}
\rho_x ({\bf  r}, {\bf  r'}) &=& 3 \rho_0 (x) \frac{J_1 \left( k_f (x) \vert {\bf  r} - {\bf  r'} \vert \right)}{\left( k_f (x) \vert {\bf  r} - {\bf  r'} \vert \right)} \nonumber \\
&&\times \Theta \left(x-\frac{ \vert {\bf  r} + {\bf  r'} \vert }{2} \right),
\label{denrr}
\end{eqnarray}
where J$_1$ is said to be the spherical Bessel function kind of first order and $k_{f}$ is the Fermi momentum of nucleons inside the {\it  Flucton} having radius $x$ and  $k_f (x)=(3\pi^2/2\rho_0(x))^{1/3} =\gamma/x$, ($\gamma\approx 1.52A^{1/3}$). The Wigner distribution function of the OBDM of Eq. (\ref{denrr}) is given by,
\begin{eqnarray}
W ({\bf  r}, {\bf  k}) =  \int_0^{\infty} dx\, \vert F(x) \vert^2\, W_x ({\bf  r}, {\bf  k}).
\label{wing}
\end{eqnarray}
Here, $W_x ({\bf  r}, {\bf  k})=\frac{4}{8\pi^3}\Theta (x-\vert {\bf  r} \vert)\Theta (k_F(x)-\vert {\bf  k} \vert)$. The density $\rho$ (r) in terms of weight function within the CDFM access is:
\begin{eqnarray}
\rho (r) &=& \int d{\bf  k} W ({\bf  r}, {\bf  k}) \nonumber \\
&& = \int_0^{\infty} dx\, \vert F(x) \vert^2\, \frac{3A}{4\pi x^3} \Theta(x-\vert{\bf  r} \vert),
\label{rhor}
\end{eqnarray}
which is normalized to A, i.e.,  $\int \rho ({\bf  r})d{\bf  r} = A$. In the $\delta$-function limit, the Hill-Wheeler integral equation, that is the differential equation for the WF in the generator coordinate is retrieved \cite{antozphys1980}. The $|F(x)|^2$ for a provided density $\rho$ (r) is described as here
\begin{equation}
|F(x)|^2 = - \left (\frac{1}{\rho_0 (x)} \frac{d\rho (r)}{dr}\right)_{r=x},
\label{wfn}
\end{equation}
with $\int_0^{\infty} dx \vert F(x) \vert^2 =1$. A detailed genealogy can be found in Refs. \cite{bhu18,anto4,anto2,gad11,gad12}. The finite nuclear symmetry energy $S^{A}$ is calculated by weighting the corresponding quantity for infinite NM within the CDFM, as given below \cite{anto4,gad11,gad12,fuch95,anto17}
\begin{eqnarray}
S^{A}= \int_0^{\infty} dx\, \vert F(x) \vert^2\, S^{NM} (\rho (x)) ,
\label{s0}
\end{eqnarray}
The $S^{A}$ in  Eq. ($\ref{s0}$) are the surface weighted average of the corresponding NM quantity at local density for finite nuclei. Again, it is worth manifesting that in CDFM, the nuclear/nucleonic density does not follow any sharp edge surface; thus, the diffuseness parameter is also not neglected. The fluctuation of the flucton attains infinitesimal size/dimension and is distributed over the range of nuclear density distribution from the relativistic mean-field model. Hence, there is no possibility of sloppy surface contribution from the density in terms of weight function.

\section{Results and discussions}
\label{results} \label{3pt}
The quantities such as binding energy (B.E), charge radius ($R_{ch}$), symmetry energy, and so on are taken as observable for the determination of isotopic shift in the present analysis. Firstly, we highlight the bulk properties such as binding energy, mean deviation of binding energy, nuclear charge radius, two neutron separation energy, neutron skin thickness, neutron chemical potential energy, pairing energy, symmetric energy, single-particle energy, occupation probability of the above-considered nuclei. The symmetry energy is calculated by using Br\"uckner prescription. The well-known three-point method \cite{georges2019} is used to obtain the shift over the isotopic chain for above defined observable. The three-point formula can be expressed as, 
\begin{eqnarray}
\triangle_{kn} {\cal{O}}(Z,N) \equiv \frac{1}{2} \left [{\cal{O}}(Z,N+k) - 2{\cal{O}} (Z,N)+{\cal{O}}(Z,N-k)\right ].
\label{three-point}
\end{eqnarray}
Here $\cal{O}$ stands for the observable, and $k$ = 2 corresponds to curvature/kink parameter. In our present work, we have considered different bulk quantities like (B.E), charge radius ($R_{ch}$), paring energy, separation energy,  neutron chemical potential ($\lambda_n$) as our observable.\\

\begin{table}
\caption{The mean deviation (MD) in binding energy for RMF with G3, IOPB-1, DD-ME1, and DD-ME2 parameter sets with respect to the experimental data \cite{wang2017} for Ca, Sn, Pb, and superheavy nuclei (SHN).}
\renewcommand{\tabcolsep}{0.22cm}
\renewcommand{\arraystretch}{2.0}
\begin{tabular}{c|ccccccccc}
\hline\hline 
Parameter & Ca & Sn & Pb & SHN & Mean\\
\hline
G3     & 0.06 & 1.99 & 3.30 & 11.66& 4.15\\
IOPB-1 & 0.5  & 1.51 & 2.95 & 8.12 & 3.27\\ 
DD-ME1 & 0.91 & 0.72 & -1.95& 9.42 & 2.27\\
DD-ME2 & 0.92 & 1.16 & 0.07 & 11.88& 3.51\\
\hline\hline 
\end{tabular}
\label{tab}
\end{table}
\begin{figure}
\includegraphics[width=1.0 \columnwidth]{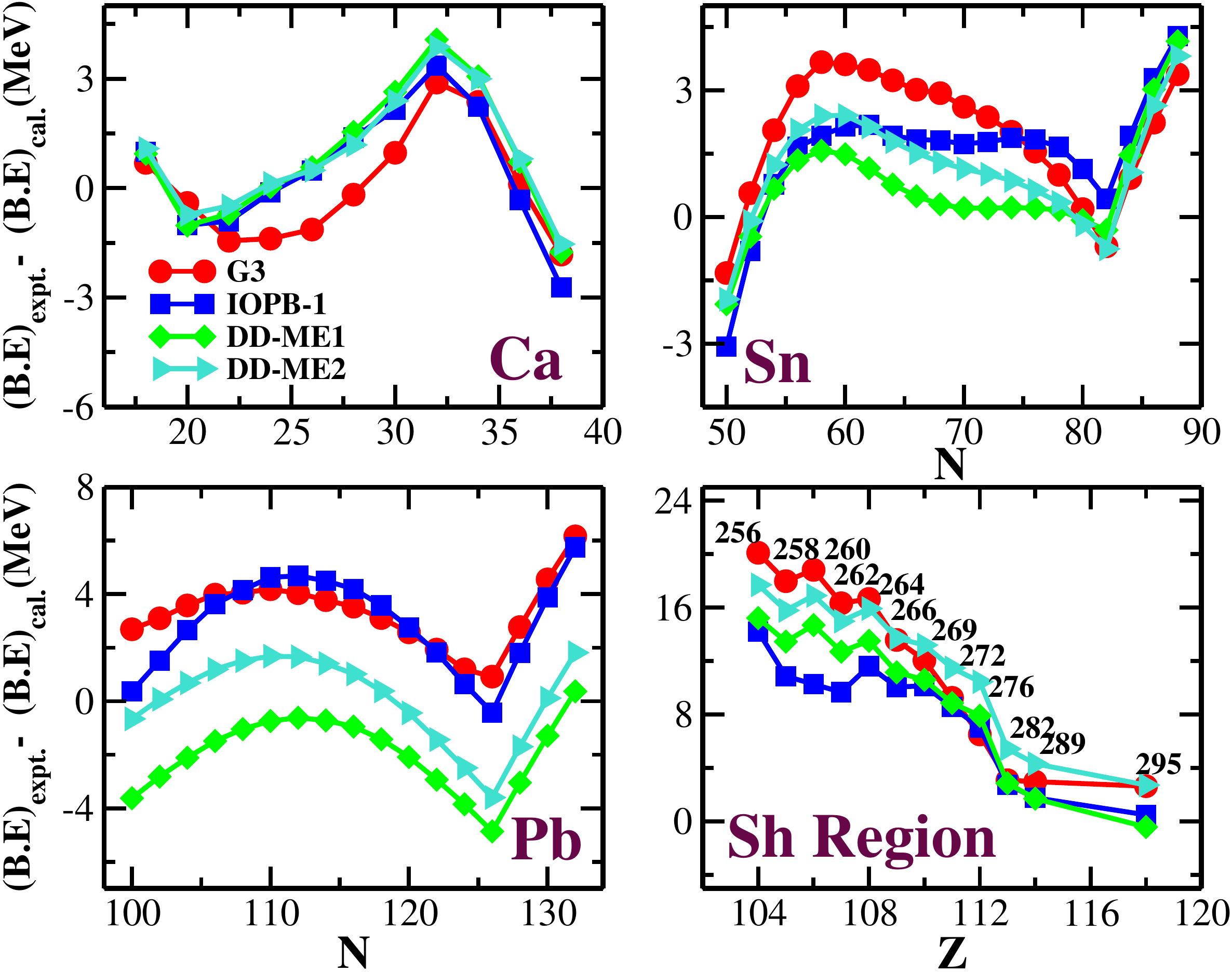}
\caption{(Color online) The binding energy (B.E) difference between the experimental data and the calculated result [$(B.E)_{expt.} - (B.E)_{cal.}$ (MeV)], is shown as a function of neutron number for Ca, Sn, Pb isotopes. Where $(B.E)_{expt.}$ and $(B.E)_{cal.}$ are the binding energies experimentally and theoretically determined respectively. In the superheavy region, we have taken the nuclei from 104 $\leq Z \leq$ 118, where experimental data available \cite{wang2017}.}
\label{fig1}
\end{figure}
\begin{figure}
\centering
\includegraphics[width=1.0 \columnwidth]{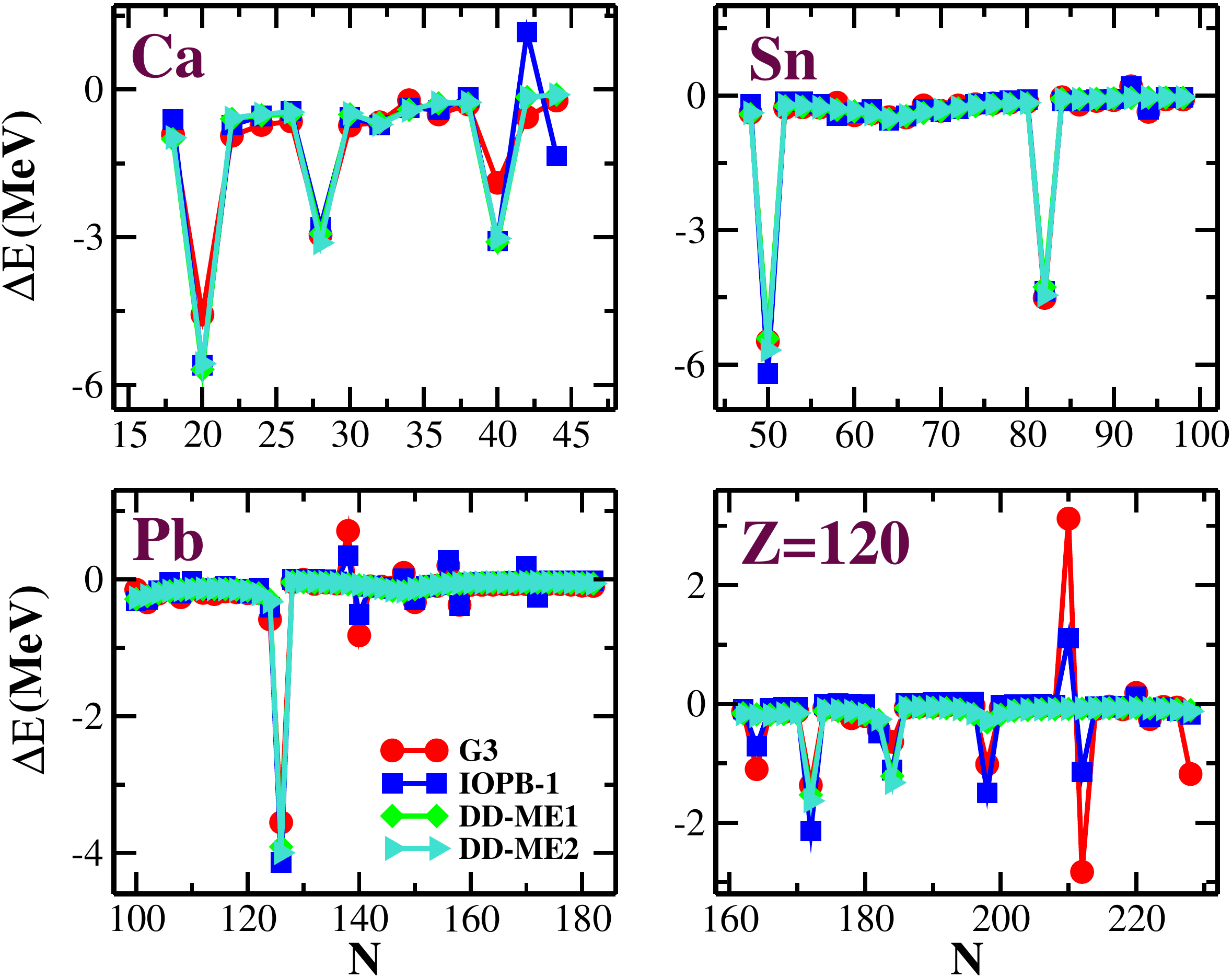}
\caption{The isotopic shift $\triangle E$ (MeV) in the B.E as a function of neutron number for Ca, Sn, Pb, and Z=120 isotopes are determined by the three-point method \cite{georges2019}. Here the peaks are emerging in downward direction.}
\label{BESHIFT}
\end{figure}
\begin{figure}
\centering
\includegraphics[width=0.5 \textwidth]{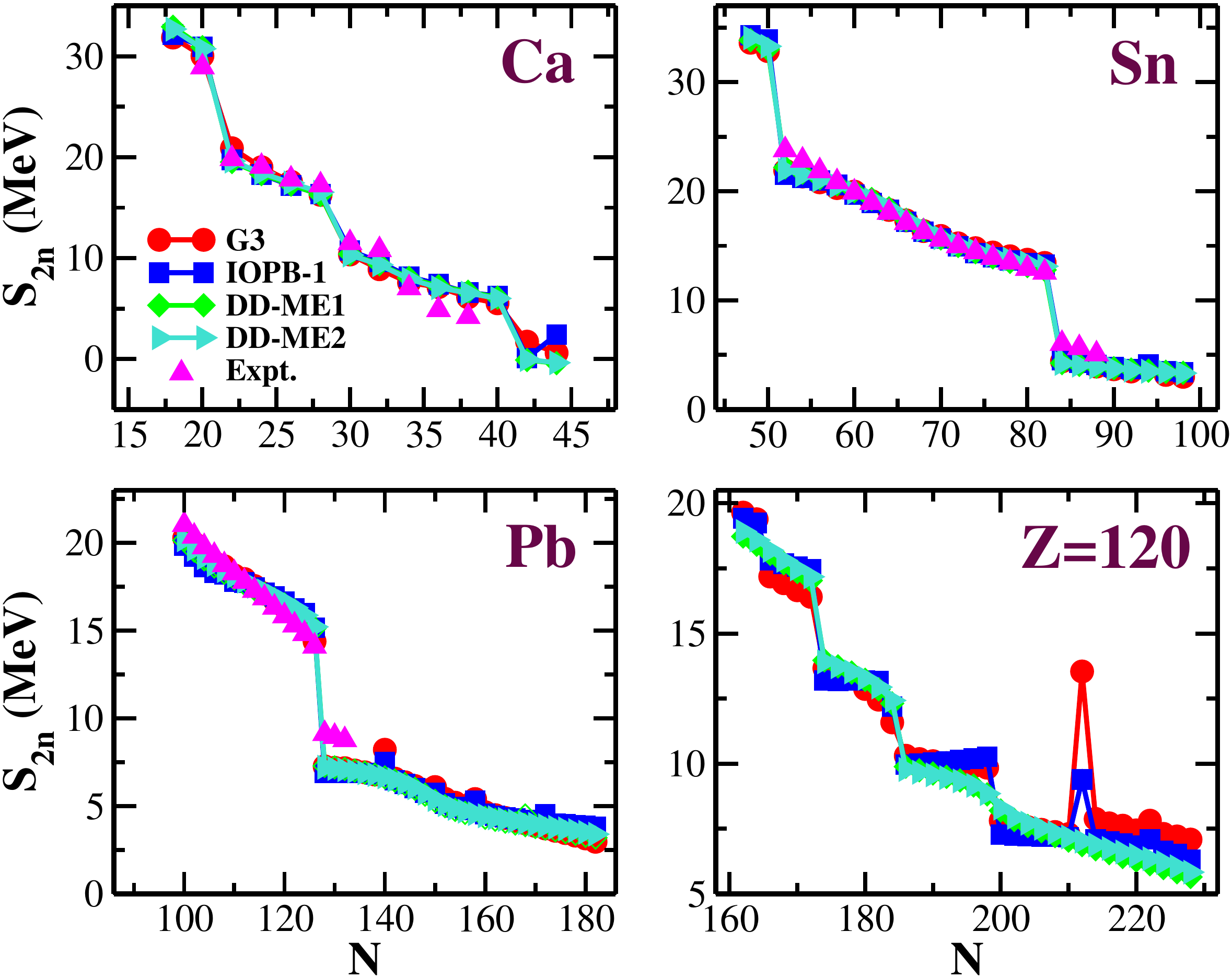}
\caption{(color online) The two neutron separation energy $S_{2n}$ (MeV) is given as a function of neutron number N for Ca, Sn, Pb, and Z=120 nuclei. The labeling in the figures shows the estimates from different parameters of RMF along with the experimental values \cite{wang2017}, wherever available.}
\label{S2N}
\end{figure}

\subsection{The binding energy and its shift}
\label{be} 
The binding energy (B.E) is a vital ground state observable, determining the shell/sub-shell closure over the isotopic chain. We have calculated the binding energy, then obtain the B.E difference between experimental data \cite{wang2017}, and calculated results, shown in Fig. \ref{fig1}. Furthermore we estimate the mean deviation (MD) of binding energy by using the formula, MD = $\frac{\sum_{i=1}^{n}(B.E)^i_{Expt.} - (B.E)^i_{RMF}}{N}$. Here, $(B.E)^i_{Expt.}$ is the experimental binding energy taken from Ref. \cite{wang2017}, and $(B.E)^i_{RMF}$ is the calculated binding energy from RMF for a specific parameter set. The estimated MDs are listed in Table \ref{tab}. The purpose of getting binding energy difference in the superheavy region is straightforward, i.e., to observe the predictions in expected shell closure Z=120 nuclei region. Finally, the isotopic shift of B.E, $\triangle E$ is investigated in fig. \ref{BESHIFT}, by three-point method of Eq. (\ref{three-point}) \cite{georges2019}. All the four-parameter sets G3 \cite{kumar17}, IOPB-1 \cite{kumar18}, DD-ME1 \cite{pring2002} and DD-ME2 \cite{pring2005}, are cooperating in nature with the empirical data. The shifts of binding energy are shown by producing peaks (downwards) at some neutron numbers, indicate the shell/sub-shell closure. The peaks follow at (N=20 and 28), (N=50 and 82), N=126, and (N=172 and 184) for Ca, Sn, Pb, and celebrity Z=120 nuclei. And also, there is a signature of sub-shell closure for N=40 in Ca nuclei. \\

\subsection{The two neutron separation energy}
\label{s2n}
The two neutron separation energy $S_{2n}$ can be obtained by using the relation, $S_{2n} (N,Z)$= B.E(N,Z) - B.E(N-2,Z). It is worth mention that the $B.E's$ for both the nuclei are calculated from RMF for G3, IOPB-1, DD-ME1, and DD-ME2 parameter sets. The experimental binding energies from Ref. \cite{wang2017} are used in the $S_{2n}$ formula to obtain the corresponding estimates and are compared in Fig. \ref{S2N} for Ca, Sn, Pb, and Z=120 nuclei. One can notice from the figure, $S_{2n}$ decreases smoothly with an increase in neutron number except for some discontinuities (i.e., kinks) at a few neutron numbers. For example, the kinks are observed at N = 20, 28, and 40 in the isotopic chain of Ca nuclei. In the case of Sn nuclei, it is found at N = 50 and 82. Similarly N = 126 for Pb and N = 172, and 184 for Z = 120 nuclei. The kink in the S$_{2n}$ energy over the isotopic chain of a specific nucleus indicates the shell/sub-shell closure of the nuclei.  It is worth here of mentioning that in energy terminology, the energy requires to remove two neutrons from a nucleus with $N_{shell} \pm 2$ is much less than the nucleus with $N_{shell}$, which appear as breaking (kink) in the regular pattern of the separation energy.   The trend we got here is so satisfactory and consistent with the available experimental data \cite{wang2017}. \\
\begin{figure}
\centering
\includegraphics[width=1.0 \columnwidth]{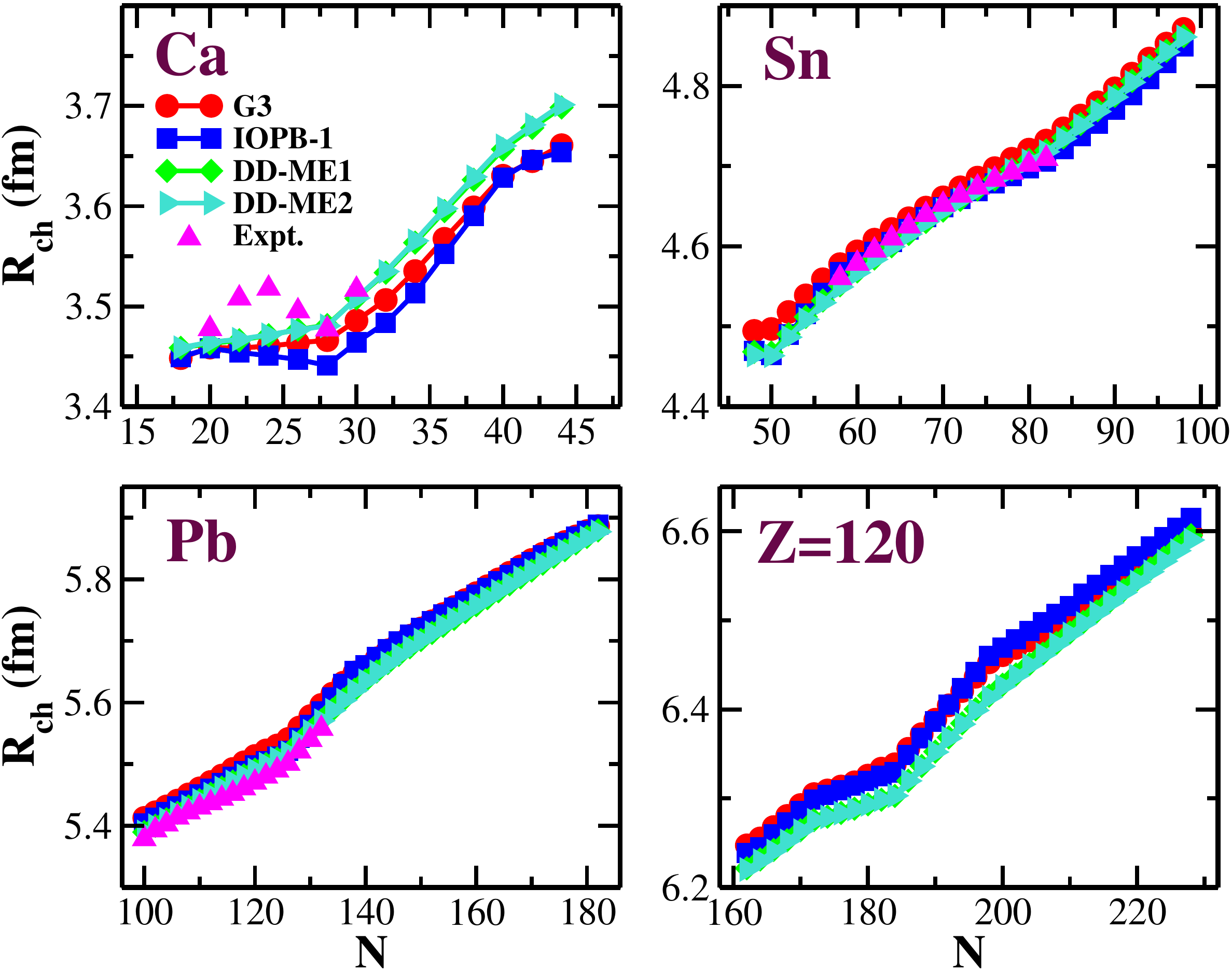}
\caption{(color online) The RMF charge radius R$_{ch}$ in $fm$ for G3, IOPB-I, DD-ME1, and DD-ME2 parameter sets as a function of neutron number for Ca, Sn, Pb and Z = 120 nuclei. The experimental data \cite{angeli} are given for comparison, wherever available.}
\label{RCH}
\end{figure}
\begin{figure}
\centering
\includegraphics[width=0.5\textwidth]{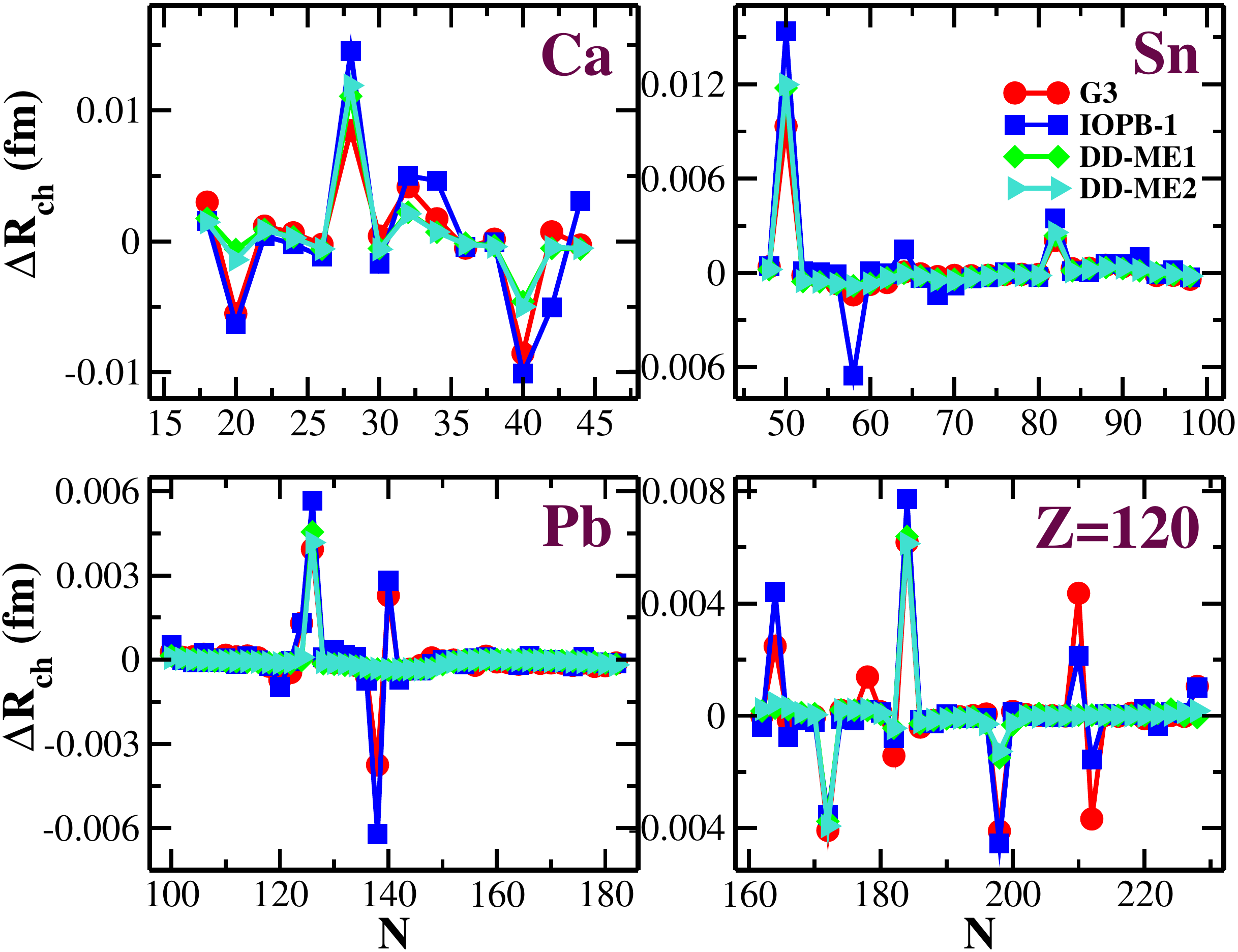}
\caption{(color online) The isotopic shift of charge radius $\Delta R_{ch}$  in $fm$ for nuclei Ca, Sn, Pb, and Z = 120 are displayed as a function of neutron number using the three-point formula of Ref.  \cite{georges2019}  along with the experimental data \cite{angeli}.}
\label{RCHSHIFT}
\end{figure}
\begin{figure}
\centering
\includegraphics[width=0.5\textwidth]{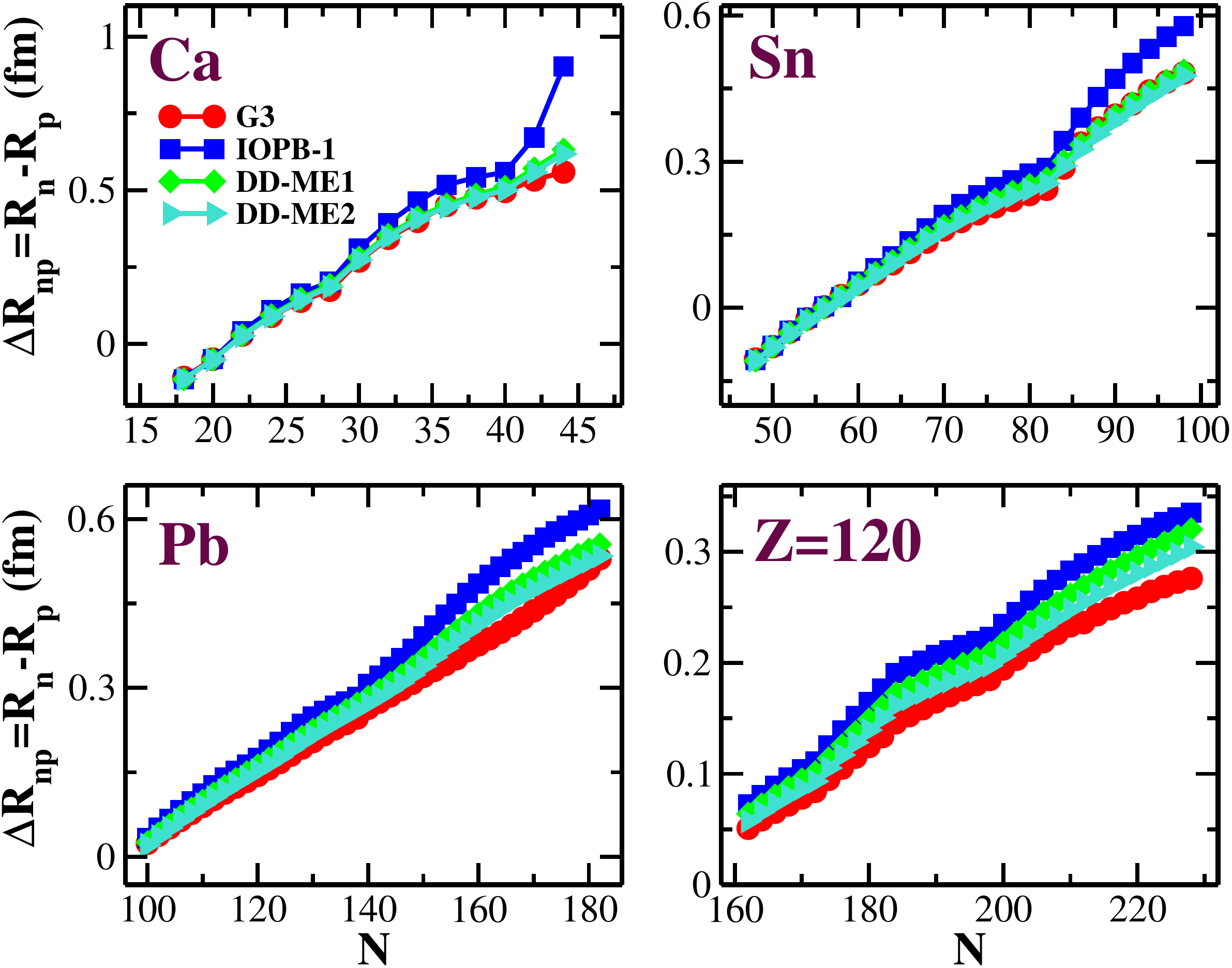}
\caption{(color online) The neutron skin thickness $\triangle{R_{np}}$ in $fm$ as a function of neutron number for Ca, Sn, Pb and Z = 120 isotopes using G3, IOPB-I, DD-ME1, and DD-ME2 parameter sets.}
\label{NST}
\end{figure}

\subsection{Isotopic shift and Skin thickness}
\label{rch} 
We have displayed the charge distribution radius R$_{ch}$ and its respective shift $\Delta R_{ch}$ as a function of neutron number for Ca, Sn, Pb, and Z = 120 isotopes. The R$_{ch}$, and  $\Delta R_{ch}$  for G3, IOPB-I, DD-ME1, and DD-ME2 are shown in Fig. \ref{RCH}, and \ref{RCHSHIFT}, respectively. There is a comprehensive discussion of the isotopic shift in the charge radius of Pb nuclei at N =126 \cite{riosprl} and reference therein. Recently, Georges {\it et al.} \cite{georges2019} observed the kink in Sn isotopes at N = 82. Hence, the isotopic shift in the charge radius can be treated as one of the key parameters to determine the magicity or island of stability in the unknown region and vice-versa. From the figure, one can find kink at N = 28, 82, 126, and 184 for Ca, Sn, Pb, and Z = 120, respectively. The experimental data \cite{angeli} for the charge radii and corresponding shift, wherever available, in good agreement with the present calculation for all the parameter sets, which shows the force independent of the results obtained. We have performed the isotopic shift calculation using the three-point formula instead of concentrating the magic neutron to justify the results from Fig. \ref{RCHSHIFT}. From the figure, one can find the peaks at the magic numbers N = 20, 28, and 40 for Ca isotopes, N=50, and 82 for Sn isotopes, N=126 for Pb. For Z = 120, the peak appears at N = 172, 184, along with two highly neutron-rich isotopes with neutron numbers N = 198 and 212 for all the force parameter sets.  Hence, more systematic studies for the magicity of these two neutron-rich isotopes, namely, $^{318,332}$120, are highly welcome. Comparing the magnitude of peaks that appear for Z = 120, we find the large magnitude corresponding to the neutron numbers N = 172 and 184. Again it confirms the shell/sub-shell closure of $^{304}$120, which is predicted to be the next double magic nuclei in the superheavy valley \cite{sobi66,meld67,nils69,bhuy12,adam09,paty91,myer66,hoff04}.   \\

In Fig. \ref{NST}, we have presented the neutron skin thickness $\triangle R_{np}= R_n – R_p$ as a function of neutron number, where $R_n$ and $R_p$ are the neutron and proton rms radii. The $\triangle R_{np}$ is an important quantity connected with the surface properties of the nucleus in terms of iso-spin asymmetry. It directly relates to the nuclear equation of state (EoS), which controls the neutron star structure and other astrophysical objects \cite{bigapple}. We have shown the neutron skin thickness for Ca, Sn, Pb, and  Z=120 isotopic series for G3, IOPB-I, DD-ME1, and DD-ME2 parameter sets. As we can see, the neutron skin thickness increases with neutron number, i.e., the presence of more neutrons enhanced the nuclear radius of the nucleus. The neutron skin thickness in the nucleus is formed by combining impacts of volume and surface term. In other words, the volume part gives an idea about an increase in the local mean-field of the surface of the neutron to the proton individual. Similarly, the surface part in which the neutron's surface width increases with mass number A in an isotopic chain. In general, both the contributions increase with neutron excess N-Z \cite{vinas2012} in an isotopic series. The linear increase in neutron skin thickness indicates the surface/volume saturation, increasing symmetric energy. We also observed the same signature for the neutron number N = 20, 28, 50, 82, 126, 172, 184 for Ca, Sn, Pb, and Z = 120. \\
\begin{figure}
\centering
\includegraphics[width=0.5\textwidth]{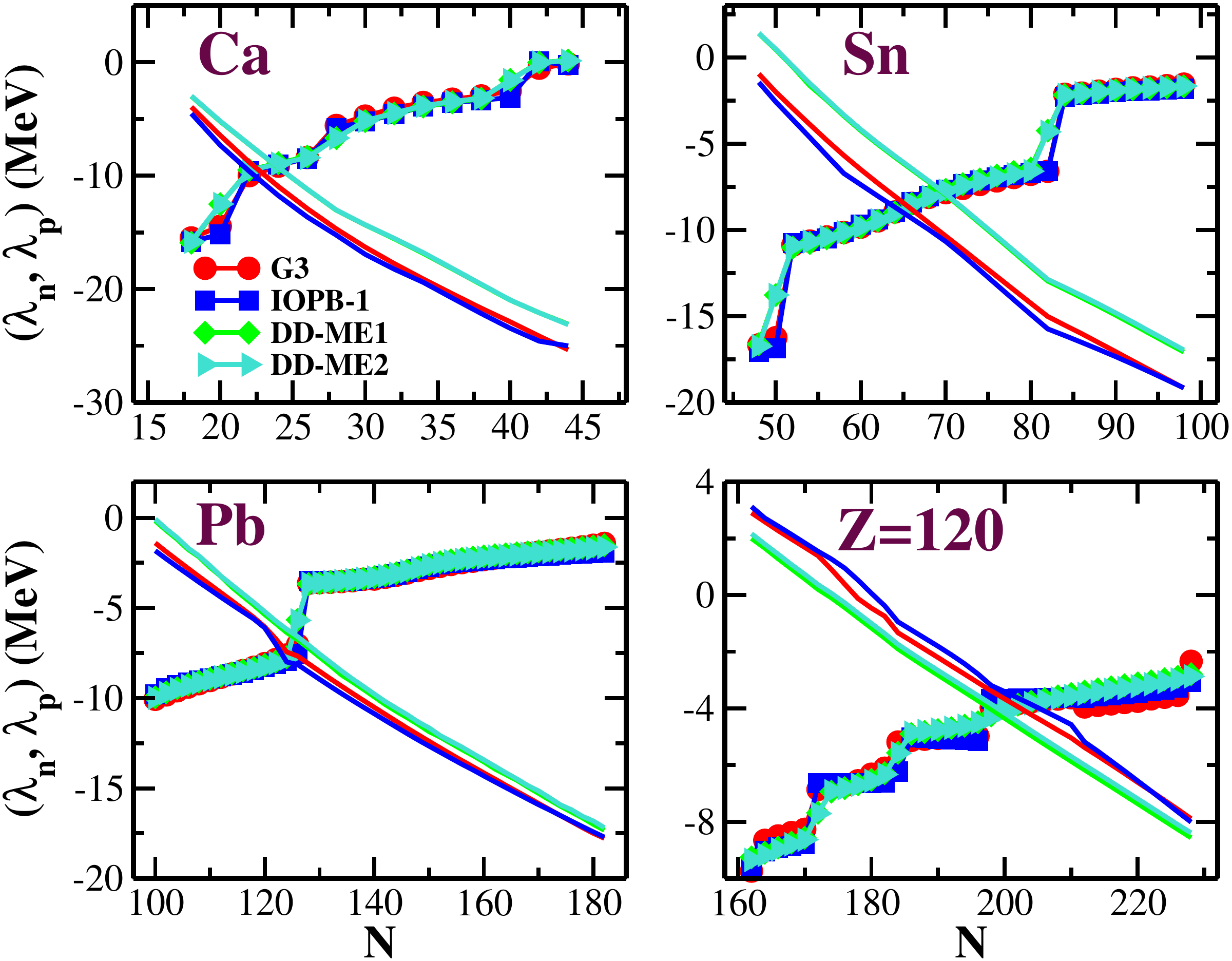}
\caption{(color online) The chemical potential energy of neutron $\lambda_n$ (MeV) and proton $\lambda_p$ (MeV) are shown for Ca, Sn, Pb, and Z=120 isotopic series as a function of neutron number. Symbols such as circle, square, diamond and triangle right are used for $\lambda_n$ (MeV) with G3 (red), IOPB-1 (blue), DD-ME1 (green) and DD-ME2 (turquoise) parameter sets, and the lines without the symbols represent $\lambda_p$ (MeV).}
\label{LAMBDAN}
\end{figure}
\begin{figure}
\centering
\includegraphics[width=1.05\columnwidth]{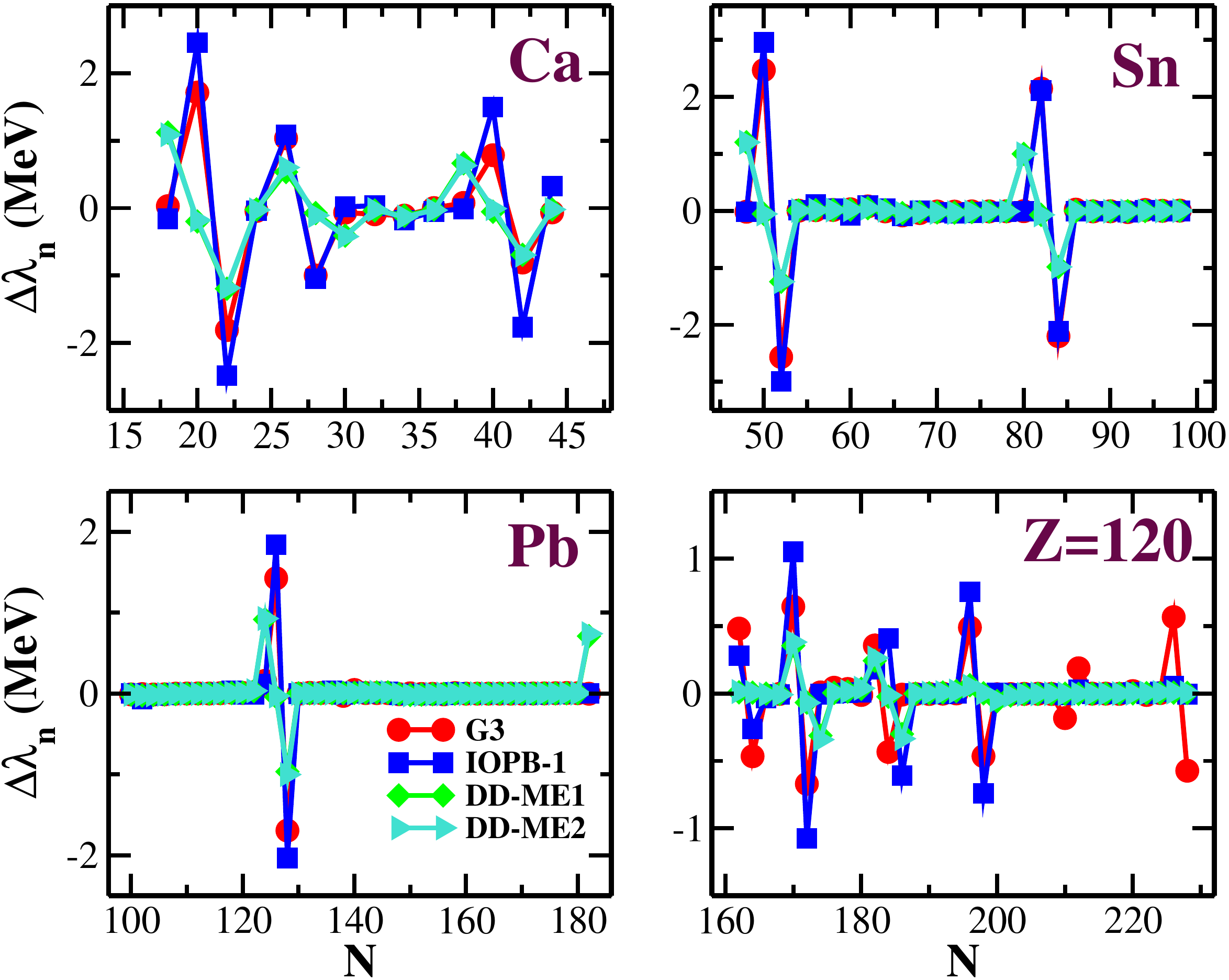}
\caption{(color online) The isotopic shift of neutron chemical potential energy $\Delta \lambda_n$ (MeV) is presented here as a function of neutron number for Ca, Sn, Pb and Z=120 nuclei.}
\label{CHEMPOTSHIFT}
\end{figure}
\subsection{Chemical potential energy}
\label{chep} 
Chemical potential energy plays an essential role in the stability of the nucleus related to shell closure  \cite{baldoplb,tsahoo} and nuclear symmetry energy \cite{baldoinis}. One can argue that it is energy desired to add a new nucleon to the nucleonic system. Further, the quantity of energy per particle increases as they are adding on. The chemical potential for proton ($\lambda_p$) and neutron ($\lambda_n$) are calculated within RMF for G3, IOPB-I, DD-ME1, and DD-ME2 parameters sets. Using the three-point formula, we estimate the shift in the $\lambda_n$ for the isotopic chain of   Ca-, Sn-, Pb-, and Z = 120 nuclei. The obtained results for the chemical potentials and the shift in the neutron chemical potential ($\Delta \lambda_n$) are shown in Fig. \ref{LAMBDAN}, and \ref{CHEMPOTSHIFT}, respectively. From the figure, we notice that the $\lambda_p$ decreases linearly with the neutron number. In the $\lambda_n$, there is a sudden jump at a few neutron numbers followed by a linear growth curve. A similar trend with a significant peak was observed in the shift of neutron chemical potential for all isotopic chains.  Here, we again confirm the traditional neutron magic at 20, 28, 50, 82, 126 for Ca-, Sn-, and Pb- isotopes. Beside these, we find the shell/sub-shell closure at N = 40 for neutron-rich ($^{60}$Ca) and N = 184 for superheavy $^{304}$120 nuclei. \\
\begin{figure}
\centering
\includegraphics[width=1.0\columnwidth]{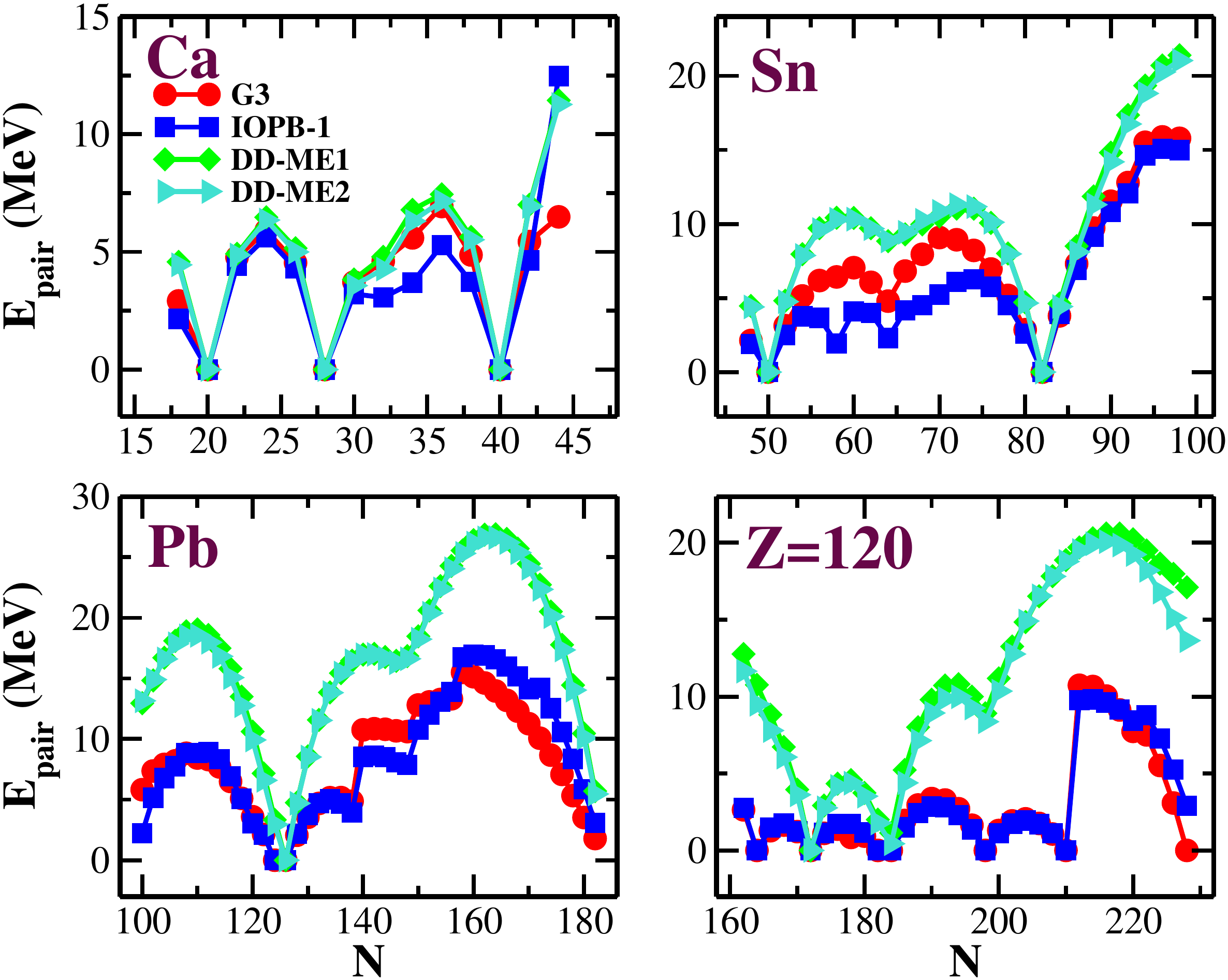}
\caption{(color online) The total pairing energy $E_{pair}$ (MeV) as a function of neutron number for Ca, Sn, Pb and Z = 120 isotopes within relativistic mean-field model for G3, IOPB-I, DD-ME1, and DD-ME2 parameter sets.}
\label{PAIRING}
\end{figure}
\begin{figure}
\centering
\includegraphics[width=1.0 \columnwidth]{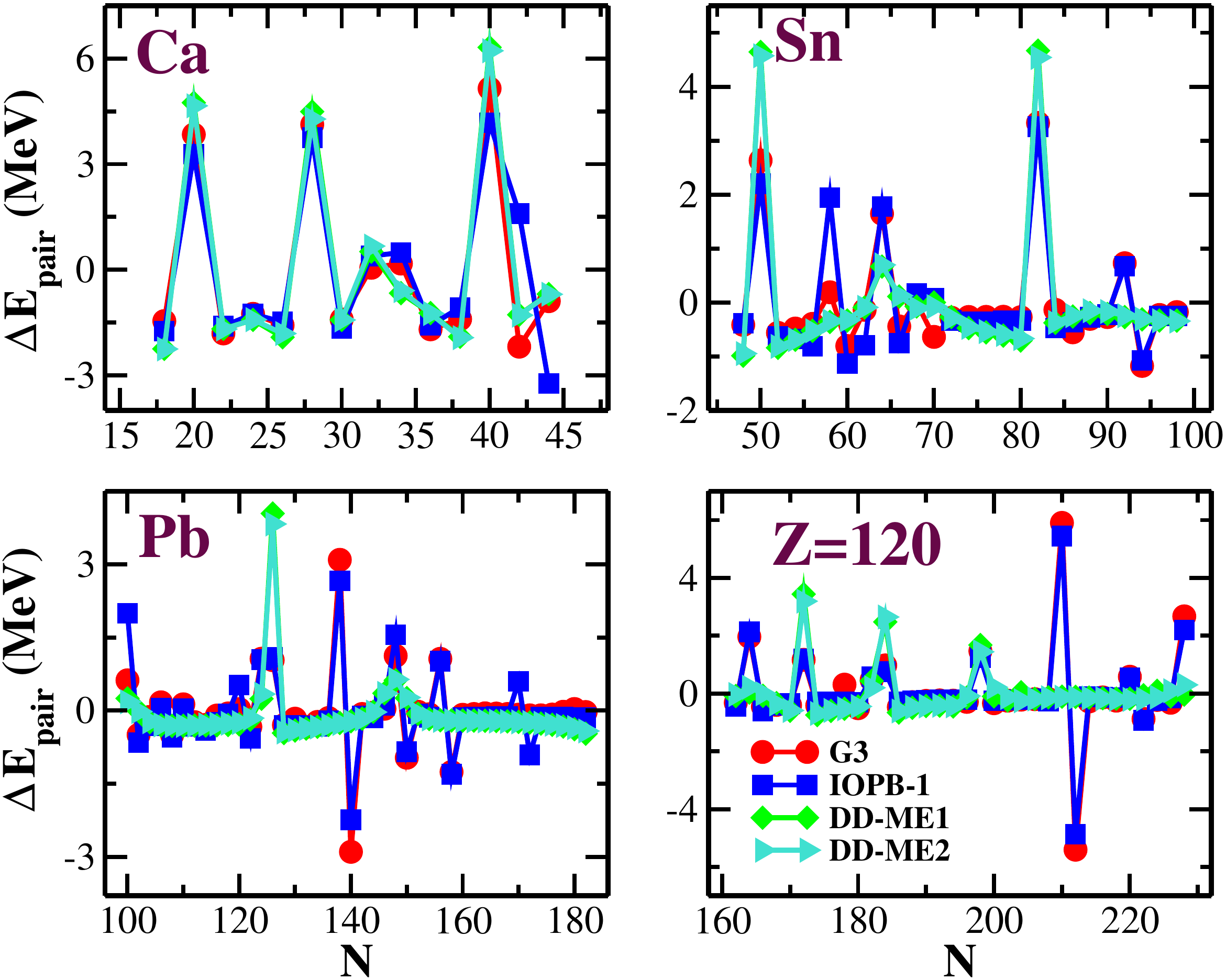}
\caption{(color online) The pairing energy shift $\Delta E_{pair}$ as a function of neutron number for Ca, Sn, Pb and Z=120 nuclei.}
\label{PSHIFT}
\end{figure}

\subsection{Pairing energy}
\label{pae} 
The pairing energy E$_{pair}$ is associated with the valence nucleon (s) and one of the essential parameters in determining the stability of the nucleus. For example, the nucleon pairs with odd numbers of N and Z have lower binding energy than the even-even system. The pairing energies from RMF with G3, IOPB-I, DD-ME1, and DD-ME2 for Ca, Sn, Pb, and Z = 120 isotopic chains are presented in Fig. \ref{PAIRING}.  Further, we have obtained the shift of the pairing energy $\Delta E_{pair}$ using the three-point formula as mentioned above, and the results are shown in Fig.\ref{PSHIFT}. One can notice from the figure that pairing energy produces the dip (s) and/or approaching zero in magnitude at a particular neutron number (s). The pairing energy approaches to zero value, i.e., the point where pairing energy collapses, which shows the shell/sub-shell closure over the isotopic chain \cite{bhumpla} and references therein. The same trends in opposite directions can also be observed from the pairing energy shift, which visualizes the relative change in the paring energy concerning neighboring isotopes. Here again, it reconfirm the shell/sub-shell closures at neutron numbers N = 20, 28, and 40 for Ca, N=50, 82 for Sn, N = 126 for Pb, and at 172, and 184 for Z = 120. \\

\begin{figure}
\centering
\includegraphics[width=1.0 \columnwidth]{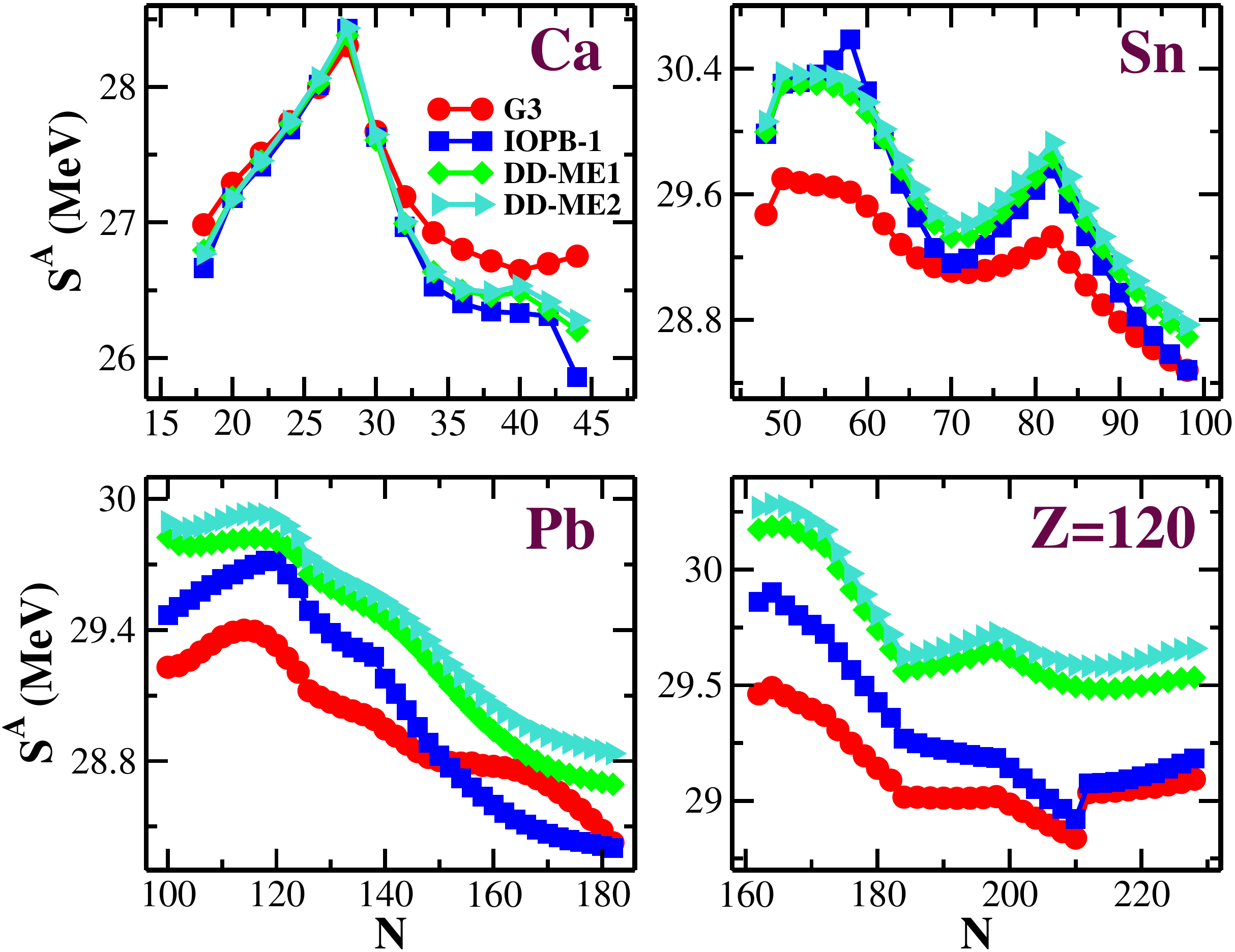}
\caption{(color online) The symmetry energy $S^A$ (MeV) is displayed as a function of neutron number for Ca, Sn, Pb and Z = 120 nuclei.}
\label{S0P0K0}
\end{figure}
\begin{figure}
\centering
\includegraphics[width=1.0 \columnwidth]{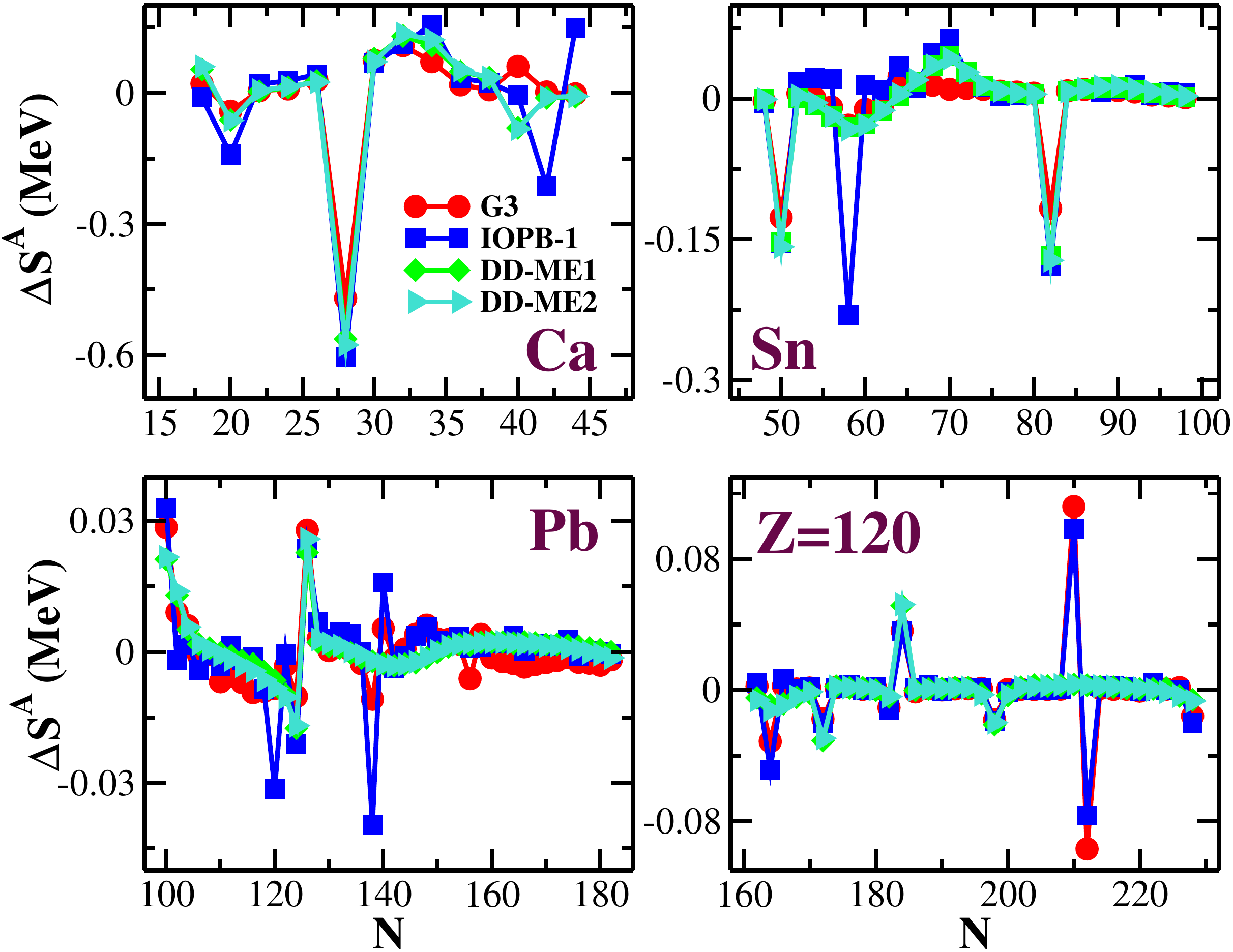}
\caption{(color online) The shift in the symmetry energy $\Delta S^A$ (MeV) over the isotopic chain for Ca, Sn, Pb and Z=120 nuclei as a function of neutron number for G3, IOPB-I, DD-ME1, AND DD-ME2 parameter sets.}
\label{S0P0K0SHIFT}
\end{figure}

\subsection{Symmetry energy of finite nuclei}
\label{syme}
There are several prescriptions to analyze nuclear surface properties of finite nuclei in terms of iso-spin and density-dependent nuclear matter quantity, such as symmetry energy and its derivatives. We have given attention to the CDFM approach of Br\"uckner's functional, and further the volume and surface components of the symmetry energy. The symmetric energy and nuclear matter density at saturation need in the Br\"uckner's functional as of Eq. (\ref{s0}) to formulate the symmetric energy of a finite nucleus. The calculated nuclear symmetric energy $S^A$ and its shift $\Delta S^A$ over the isotopic chain are shown in Figs. \ref{S0P0K0} and \ref{S0P0K0SHIFT}, respectively, as a function of the neutron number. We have presented the symmetric energy for the isotopic chain of considered doubly-magic nuclei Ca, Sn, Pb, and expected magic nuclei Z = 120 for G3, IOPB-I, DD-ME1, and DD-ME2 parameter set. Fig. \ref{S0P0K0} shows that the magnitude for different isotopic chains varies with the neutron number. We find a few peaks at shell/sub-shell closures corresponding to the neutron number N = 20, 28, 40, 50, 82, 126, 172, and 184 (see Fig. \ref{S0P0K0}). It is worth mentioning that there is no significant peak in the symmetry energy at N = 126 for $^{208}$Pb, which is even studied in some of the earlier work of Ref. \cite{quddus2020} and reference therein. Here we also find the same, but when estimating the shift over the isotopic chain by using the three-point method described in sec. (\ref{3pt}), we observed the peaks at N = 126 as shown in Fig. \ref{S0P0K0SHIFT}. Furthermore, the present work also provides and/or confirms the magicity for highly neutron-rich $^{60}$Ca and the island of stability at superheavy valley $^{304}$120. It is worth demonstrating that nuclei in the superheavy region have an excellent response to the nature of nuclear symmetry energy. Further, the appropriate information of symmetry energy from finite nuclei will be added to a wide range of nuclear phenomena, starting from the study of nuclear structure, dynamics of heavy-ion reactions to the high iso-spin asymmetry system neutron star matter. \\

\begin{figure}
\centering
\includegraphics[width=1.0\columnwidth]{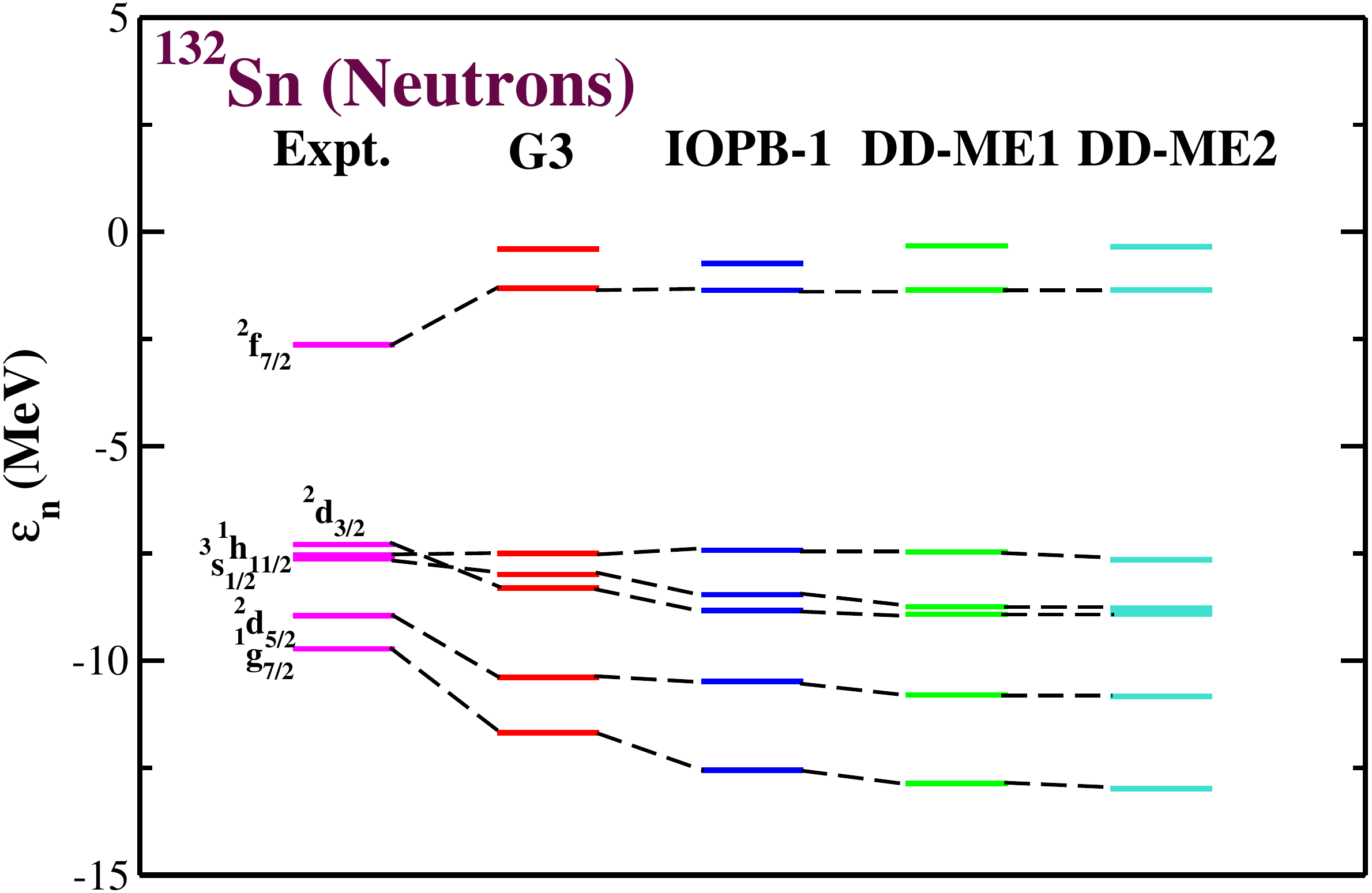}
\includegraphics[width=1.0\columnwidth]{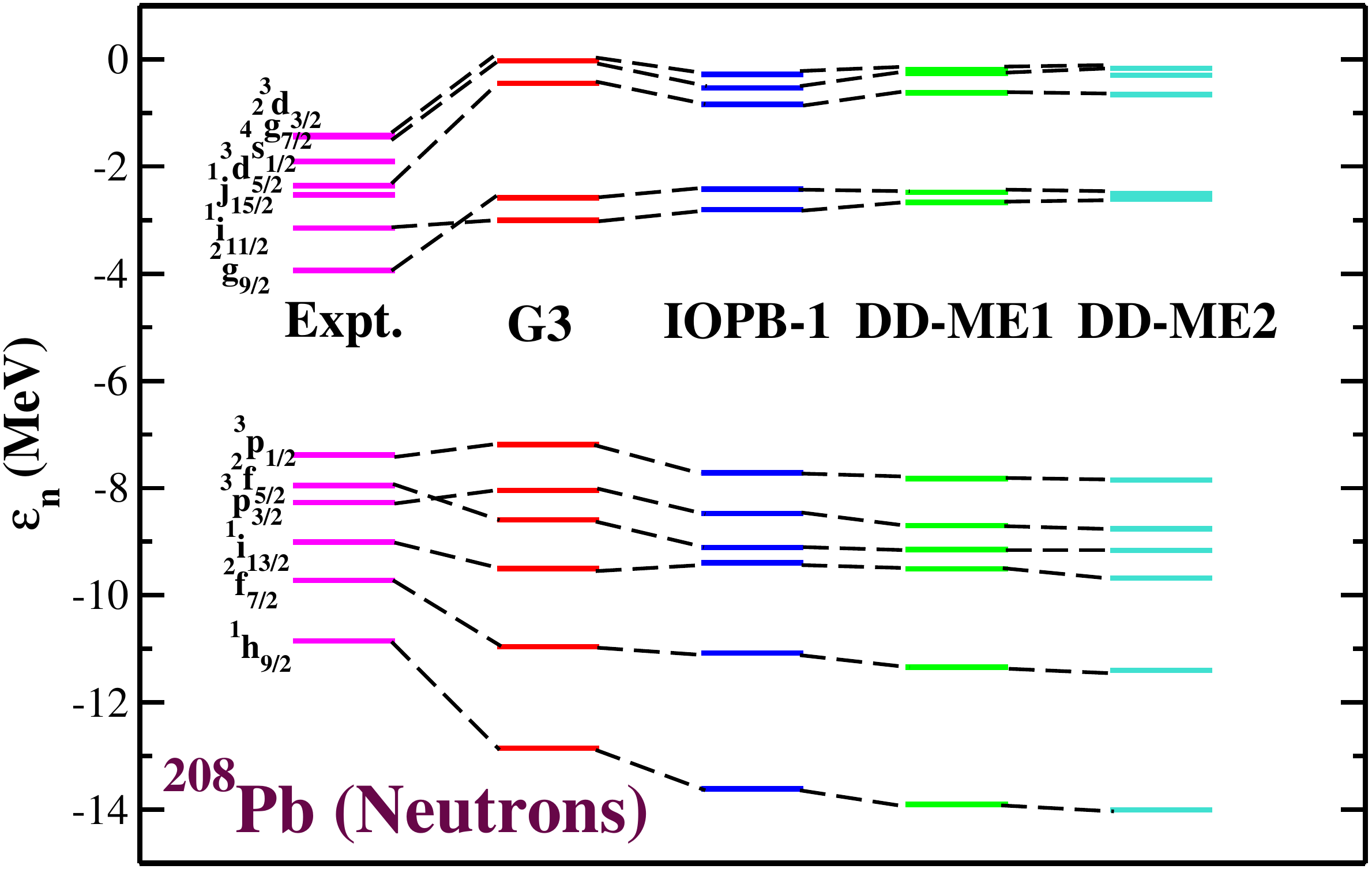}
\includegraphics[width=1.0\columnwidth]{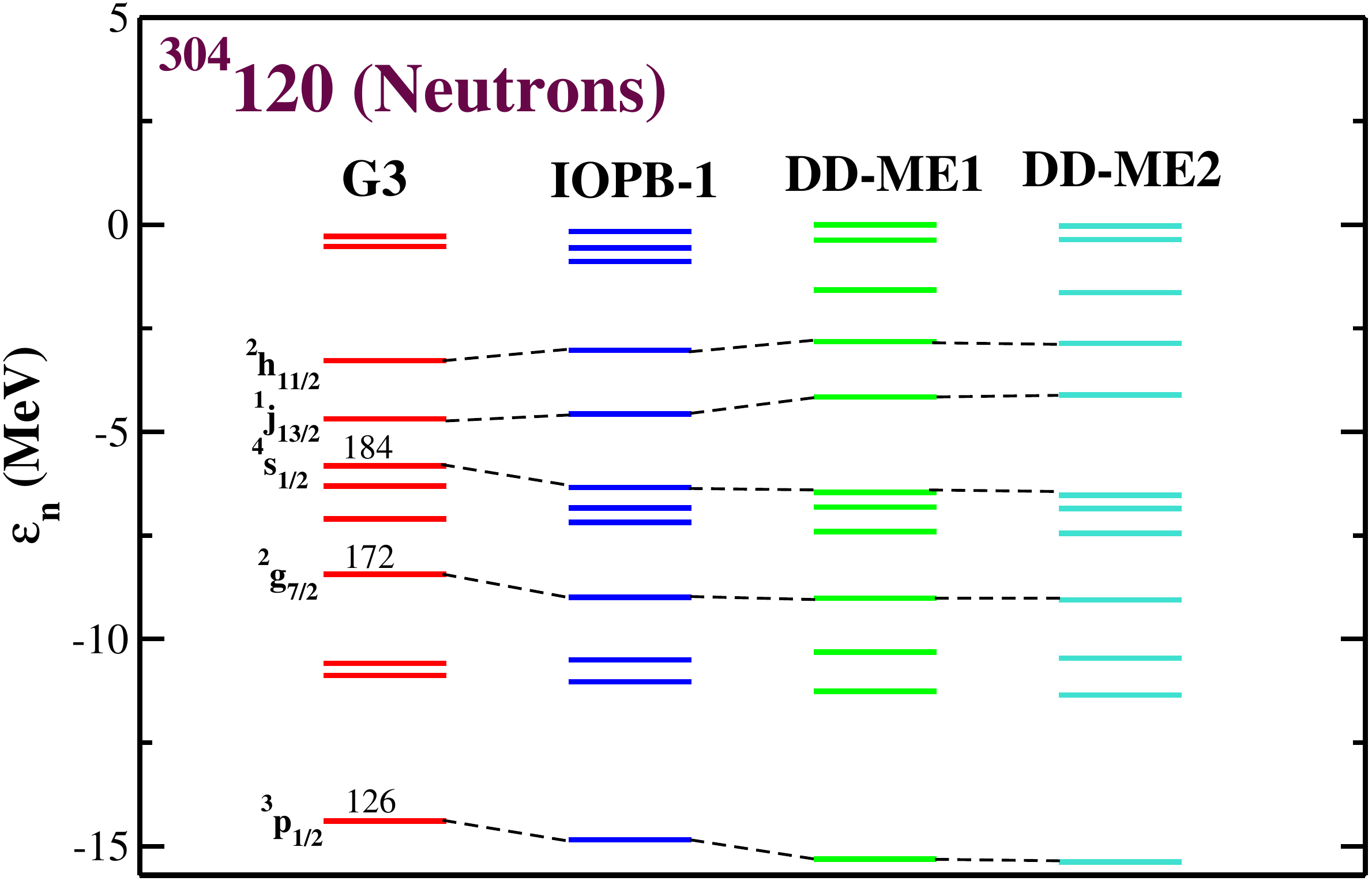}
\caption{(color online) The neutron single particle energy $\epsilon_n$ of $^{132}$Sn, $^{208}$Pb, and $^{304}$120 nucleus along with the available experimental data \cite{brink72,dudek84}. Similarly for G3, IOPB-1, DD-ME1 and DD-ME2 parameter sets the colours are red, blue, green and turquoise respectively.)}
\label{sn132n}
\end{figure}
\begin{figure}
\centering
\includegraphics[width=1.0 \columnwidth]{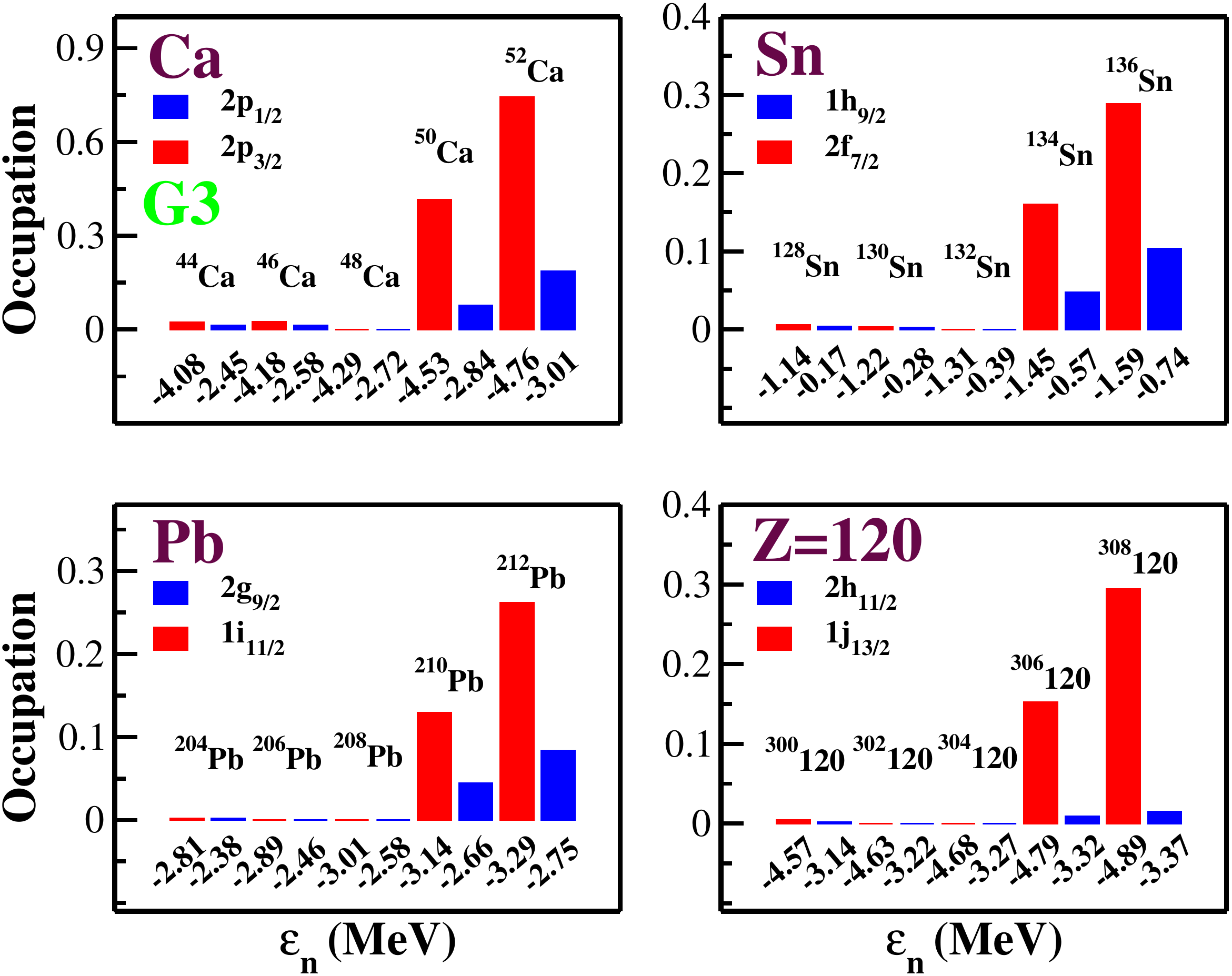}
\caption{(color online) The occupation probability of considered neutron orbitals are determined by RMF G3 parameter set and presented as the function of their single particle energies (MeV) for Ca, Sn, Pb and Z=120 nuclei.}
\label{occp}
\end{figure}

\subsection{Single particle energy and Occupation}
\label{occ} 
The single-particle energies  (SPEs) are calculated from the relativistic mean-field model for G3, IOPB-I, DD-ME1, and DD-ME2 for $^{132}$Sn, $^{208}$Pb and $^{304}$120 nuclei. The neutron SPEs ($\epsilon_n$) near the Fermi surface for $^{132}$Sn, $^{208}$Pb and $^{304}$120 along with the available empirical data \cite{brink72,dudek84} are displayed in the upper, middle, and lower panel of Fig. \ref{sn132n}, respectively. In Fig. \ref{occp} we have shown the occupation probabilities of neutron orbitals corresponding to the neutron numbers N = 28, 82, 126, and 184, for Ca, Sn, Pb, and Z = 120 nuclei for the G3 RMF parameter set. Similar trends can be expected for IOPB-I, DD-ME1, and DD-ME2 parameter sets, which are not given to retaining the figure's clarity. We have observed the filling of the valence orbital following the neutron magic, i.e., N = 28, 82, 126, and 184 for Ca, Sn, Pb, and Z =120, respectively. For example, the 2$p_{1/2}$ \& 2$p_{3/2}$ is taken as the reference orbitals for $^{48}$Ca isotope for observation. Similarly,  1$h_{9/2}$ \& 2$f_{7/2}$, 2$g_{9/2}$ \& 1$i_{11/2}$, and 2$h_{11/2}$ \& 1$j_{13/2}$ are the reference orbitals of observation for $^{132}$Sn, $^{208}$Pb and $^{304}$120, respectively. The occupation probability or the filling of the orbitals play a great role in explaining the isotopic shift in the charge radii over the isotopic chain \cite{riosepj,georges2019,bhujpg2021}. Following the isotopic shift at N = 126 can explain through the earlier filling of 1$i_{11/2}$ orbitals than 2$g_{9/2}$. From a nuclear matter perspective, one can see that the nuclear symmetry energy acts to increase the overlap between neighboring orbital and the overall density. The effect will be increased if the overlap between those wave functions is maximum \cite{riosprl,bhujpg2021}.

A similar trend can find for their corresponding orbitals for the Ca-, Sn-, and Z= 120 isotopes. By observing the occupation number in parallel with the neutron single-particle energies (see Fig. \ref{occp}) for the G3 parameter set, we notice the occupation probability of $2p_{3/2}$ is filling earlier than $2p_{1/2}$ for Ca. In the case of Sn- isotopes, the $2f_{7/2}$ has more occupation than $1h_{9/2}$ neutron orbital. Following the trends, we find $1j_{13/2}$ is more occupied than $2h_{11/2}$, associated with the predicted to be the neutron magic N = 184. From the above analysis, it is evident that the neutron orbitals corresponding to the higher energy are mostly occupied, as they are more deeply bound. From the above analysis, one can connect the occupation probability with the trend of magicity and shift in various observables corresponding to the neutron magic and vice-versa. \\

\section{Summary}
\label{summary}
Although we were very much aware of the isotopic shift of charge radius in Pb (N = 126) and Sn (N = 82), here we observe the kinks and/or peaks in the various observable over the isotopic chain of Ca, Sn, Pb, and Z = 120 nuclei. We observe that the shifts in an isotopic chain for a particular atomic nucleus can be correlated with the shell/sub-shell closures. The ground state bulk properties such as binding energy, root-mean-square radius, pairing energy, nuclear density distributions, and single-particle energies are calculated for the isotopic chain of Ca, Sn, Pb, and Z = 120 nuclei. The relativistic mean-field with G3, IOPB-I, DD-ME1, and DD-ME2 parameter sets are used in the present analysis. To see the applicability of the RMF model, we estimate the binding energy difference and mean deviation to the experimental data and found less than 1\% of the variation in magnitude for considered isotopic chains. The isotopic shift of charge radius and neutron skin thickness is estimated for all the isotopic chains. We also obtain the shift in the binding energy, pairing energy, and chemical potential for all the isotopic chains using the three-point formula. The two neutron separation energy is estimated using the ground state binding energy for each isotope. The symmetry energy, which is one of the essential parameters to define the magicity of the nuclei far from the $\beta$-stable region of the nuclear chart \cite{abdul120,quddus2020,bhu18,gad11,gad12,abdulsn}, is also calculated for all isotopic chains. It is worth mentioning that the Coherent Density Fluctuation Model is used to determine the symmetry energy within the framework of the relativistic mean-field formalism \cite{bhu18} and reference therein. Furthermore, the shift in the symmetry energy over the isotopic chain is estimated for all the parameter sets. \\

We reconfirm the known neutron magics N = 20, and 28 for Ca-, N = 50, and 82 for Sn-, and N = 126 for Pb- isotopic chain from the above analysis. We establish the shell/sub-shell closure at N = 40 for neutron-rich $^{60}$Ca, and re-authenticate N = 184 for superheavy $^{304}$120. We also noticed the signature of magicity at N = 198 \& 212 for Z = 120 isotopic chains. The single-particle energy levels for proton and neutron are also estimated for all the parameter sets within relativistic mean-field formalism. To reduce the number of figures, we have shown the neutron single-particle energies for $^{132}$Sn, $^{208}$Pb, and $^{304}$120 along with the available experimental data. The corresponding occupation probabilities for 2$p_{1/2}$ \& 2$p_{3/2}$, 1$h_{9/2}$ \& 2$f_{7/2}$, 2$g_{9/2}$ \& 1$i_{11/2}$, and 2$h_{11/2}$ \& 1$j_{13/2}$ are given as the reference orbitals of observation for $^{48}$Ca, $^{132}$Sn, $^{208}$Pb and $^{304}$120, respectively. The correlation is established between occupation probability and the magicity of a nucleus in an isotopic chain for a particular atomic nucleus. Furthermore, we observed that the shift of various structural observables could be considered the key parameter for determining the shell/sub-shell closures over an isotopic chain and vice-versa. \\

\noindent 
{\bf  Acknowledgement:} One of the authors (JAP) is thankful to the Institute of Physics, Bhubaneswar, for providing computer facilities during the work. SERB partly reinforces this work, Department of Science and Technology, Govt. of India, Project No. CRG/2019/002691. MB acknowledges the support from FOSTECT Project No. FOSTECT.2019B.04, FAPESP Project No. 2017/05660-0, and the CNPq - Brasil. \\



\end{document}